\documentclass[12pt]{article}
\usepackage{tikz}
\usepackage{amsmath}
\usepackage{mathtools}
\usepackage{fancyhdr}
\usepackage{gensymb}
\usepackage{graphicx}
\usepackage{bbold}
\usepackage{cite}
\usetikzlibrary{decorations.pathmorphing, arrows, patterns}
\tikzset{snake it/.style={-stealth,
decoration={snake, 
    amplitude = .4mm,
    segment length = 2mm,
    post length=0.9mm},decorate}}
\usepackage{amssymb}
\usepackage{cancel}

\usepackage[a4paper, total={6in, 10in}]{geometry}

\begin{document}
\title{\Large{\textbf{Diffeomorphisms of the energy-momentum space: perturbative QED}}}
\date{}
\author{Boris Iveti\'c
\footnote{bivetic@yahoo.com}
  \\Wöhlergasse 6, 1100 Vienna, Austria \\}

\maketitle

\begin{abstract}
A perturbative formulation of quantum electrodynamics is given in terms of geometrical invariants of the energy-momentum space, whose geometry is taken to be one of a constant curvature. The construction is relevant for different classes of noncomutativity: the Snyder model and the so called GUP models. For the Snyder model it is shown that all the amplitudes are finite at every order of the perturbation expansion.  
\end{abstract}

\section{Introduction}
That the structure of spacetime on very short scales must depart from a continuum, once the effects of gravity are taken into consideration, is a long established fact \cite{me,dop}. An importaint class of deformations of the spacetime continuum, appearing in string theory and some other models of quantum gravity, is described by the noncommutative geometry (NCG) \cite{witt,wess,sw}. 
The methods of NCG, specifically Hopf algebras, have been applied in investigations of different physical models, such as canonical noncommutativity \cite{wess,jack},  kappa-Minkowski \cite{luk1,luk2} and Snyder models \cite{mel}. This approach however runs into ambiguities when one tries to formulate a dynamical theory - classical or quantum, particles or fields - for a given type of noncommutativity. This is due to the fact that the Heisenberg algebra alone does not define dynamical principles, which can be deformed together with (and independent of) the algebra itself. These inherent ambiguities are discussed in more detail in \cite{ja1}.

At the same time, the structrure of energy-momentum space corresponding to several important classes of noncommutativity, such as GUP model \cite{adv} and Snyder model \cite{sny} is very simple - they are constant curvature spaces. As powerful as the methods of dual (Hopf) algebras may be, it seems to us that from the strategical point of view, it should in those cases be more favorable - and more natural - to approach the problem from the energy-momentum space. 
We find it a remarkable fact that this approach has not been extensively utilized in the literature. Some notable early contributions are given in the papers \cite{golf1,mk0,kady0}; of the more recent studies in this direction we refer to \cite{relloc1,ac,carm,ref8}.

Our approach in this work follows up on a programm set up in the earlier works \cite{bor,bor2,bor3} and, independently, in \cite{ref8}. In short, it consists in re-writing the standard QED - the action, Feynman rules, the amplitudes etc. - entirely in terms of the geometrical invariants of the flat energy-momentum space. Once this has been achieved, it is merely an exercise in elementary geometry to generalize these to the case of non-vanishing (constant) curvature. This demand of the covariance of the theory in the momentum representation is at the same time the missing defining principle of the dynamics. We shall in this work specifically focus on the case of negative curvature, but all our results are easily modified to the case of positive curvature. 

As an introductory example of our method, consider the dispersion relation on the flat energy-momentum space, which can also be considered as the definition of the mass:

\begin{equation}\label{dr}
m^2=E^2-\vec p^2
\end{equation}
But this can be also writen in the form
\begin{equation}
m^2=d^2(p,0)
\end{equation}
where $d(p,q)$ is the geodesic distance between points $p$ and $q$ and $p=(E, \vec p)$. So that geometrically speaking, mass of a particle with a 4-momentum $p$ is the geodesic distance from that point to the origin. Once this has been established it is easy to work out an explicit expression for the mass that holds for anti-de Sitter (adS) type of the energy-momentum space. Without this constarint, one can principally choose any dispersion relation for the adS, as long as it goes to (\ref{dr}) as the curvature is set to zero.

\section{Flat energy-momentum space}
The main interest of this paper is to define quantum electrodynamics on the adS energy-momentum space. The guiding principle in the construction is the covariance of physical laws. In this respect, this section is to be considered as a in-between step towards that goal. It can however also be considered as a specific choice (out of infinitely many) of defining QED relevant for the so called GUP models\cite{adv, add}. The latter are usually introduced through the deformation of the Heisenberg algebra,

\begin{equation}
[\hat x_\mu, \hat x_\nu]=[\hat p^\mu,\hat p^\nu]=0, \ \ \ \ \ \ [\hat x_\mu, \hat p^\nu]=f_1(\alpha^2p^2)\delta_\mu^\nu+f_2(\alpha^2p^2)\eta_{\mu\rho}p^\rho p^\nu,
\end{equation}
with functions $f_1$ and $f_2$ constrained by the demand that the algebra closes. This specific choice of the dynamics is however trivial - the QED is completely equivalent to the standard, commutative case, meaning all the values for the observed cross sections, decay widths etc. are the same as in the standard case. This is a direct consequence of our demand that the dynamics depends only on the geometry, and not on the coordinates of the energy-momentum space.

\subsection{Elements of the geometry}
In this section, we use the following notation:
\begin{align*}
\pi^\mu & - \text{standard Cartesian coordinates}\\
p^\mu & - \text{general coordinates}
\end{align*}
where $\mu=0,1,2,3$, and the two sets are related through a general isotropic diffeomorphism of the form

\begin{equation}\label{flatdef}
p^\mu=g(\alpha^2 \pi^2)\pi, \ \ \ \ \ \ \pi^\mu=h(\alpha^2p^2)p^\mu.
\end{equation}
Here $\alpha$ is a "deformation" parameter and $g$ and $h$ are arbitrary, suitably well behaved functions, satisfying in particular $g(0)=h(0)=1$. We use 

\begin{equation}
\pi^2\equiv \eta_{\mu\nu}\pi^\mu\pi^\nu, \ \ \ \ \ \ p^2\equiv\eta_{\mu\nu} p^\mu p^\nu
\end{equation}
for short.\footnote{Throughout the work, we use the notation $a^2=\eta_{\mu\nu}a^\mu a^\nu$, $(ab)=\eta_{\mu\nu}a^\mu b^\nu$, even when $a^\mu$ and $b^\mu$ are not vectors of the flat space.}

From the invariance of the square of the infinitesimal distance,
\begin{equation}
ds^2=\eta_{\mu\nu}d\pi^\mu d\pi^\nu=g_{\mu\nu}dp^\mu dp^\nu,
\end{equation}
one gets the metric in general coordinates

\begin{equation}
g_{\mu\nu}(p)=h^2\eta_{\mu\nu}+4h'\alpha^2(h+h'\alpha^2p^2)\eta_{\mu\rho}p^\rho\eta_{\nu\sigma}p^\sigma,
\end{equation}
with inverse
\begin{equation}
g^{\mu\nu}(p)=\frac{1}{h^2}\eta^{\mu\nu}-\frac{4h'\alpha^2(h+h'\alpha^2p^2)}{h^2(h+2h'\alpha^2p^2)}p^\mu p^\nu,
\end{equation}
where 
\begin{equation}
h'\equiv \frac{\partial h}{\partial(\alpha^2p^2)}, \ \ \ \ g'\equiv\frac{\partial g}{\partial(\alpha^2\pi^2)} 
\end{equation}
The determinant and the invariant volume element are 
\begin{equation}
\text{det} g=-h^6(h+2h'\alpha^2p^2)^2, \ \ \ \ \ d\Omega_p= h^3(h+2h'\alpha^2p^2)d^4p.
\end{equation}
The addition rule for the momenta is\footnote{If the two undeformed momenta $\pi$ and $\sigma$ are transformed into $p$ and $s$ respectively, then $\pi+\sigma$ must transform into $p\oplus s$ - this uniqely defines the momenta addition rule.}  
\begin{equation}
(p\oplus q)^\mu=g(\alpha^2\kappa^2)\kappa^\mu, \ \ \ \ \ \kappa^\mu\equiv h(\alpha^2p^2)p^\mu+h(\alpha^2q^2)q^\mu,
\end{equation}
where it is understood to be the displacement \textit{of} the momentum $p$ \textit{by} the momentum $q$. An alternative notation \cite{golf1} that emphasizes the general noncommutativity of the momenta addition, is

\begin{equation}
\hat d(q) p^\mu=(p\oplus q)^\mu,
\end{equation}
and also 
\begin{equation}
\hat d(q) \psi(p)=\psi(p\ominus q),
\end{equation}
where $\hat d(q)$ represents an operator of displacement for the momentum $q$, and $\psi(p)$ is an arbitrary function of $p$.

The geodesic distance between the points $p$ and $q$ is 
\begin{equation}
d(p,q)=\sqrt{h^2(\alpha^2p^2)p^2-h^2(\alpha^2q^2)q^2}.
\end{equation}
The momentum (co)vector fields are
\begin{equation}\label{flatgenmom}
f_\mu\equiv\frac{1}{2}\frac{\partial d^2(p,0)}{\partial p^\mu}=h(h+2h'\alpha^2p^2)\eta_{\mu\rho}p^\rho, \ \ \ \ f^\mu=g^{\mu\nu}f_\nu=\frac{h}{h+2h'\alpha^2p^2}p^\mu .
\end{equation}
That these are indeed the generalizations of the radial vector field can be seen from their transformation properties. The radial vector field for an arbitrary set of coordinates is defined as
\begin{equation}
p_R^\mu=\frac{\partial p^\mu}{\partial \pi^\nu}\pi_R^\nu|_{\pi=hp}=\frac{h}{h+2h'\alpha^2p^2}p^\mu,
\end{equation}
and $\pi_R^\nu=\pi^\nu$, since these are just the Cartesian coordinates of the flat space, so that

\begin{equation}
p_R^\mu=f^\mu,
\end{equation}
as follows from the construction.
The vierbein is 
\begin{equation}
e^\mu_a\equiv \lim_{q\to 0} \delta_a^\nu \frac{\partial (p\oplus q)^\mu}{\partial q^\nu}=\frac{1}{h}\delta_a^\mu-\frac{2h'\alpha^2}{h(h+2h'\alpha^2p^2)}\eta_{ab}p^bp^\mu
\end{equation}
where in the derivation the following identities that follow from (\ref{flatdef}) have been used ($h=h(\alpha^2p^2)$):
\begin{equation}
g(\alpha^2h^2p^2)=1/h, \ \ \ \ \ g'(\alpha^2h^2p^2)=\frac{\partial g(\alpha^2h^2p^2)}{\partial(\alpha^2h^2p^2)}=-\frac{h'}{h^3(h+2h'\alpha^2p^2)}.
\end{equation}

\subsection{Free fermions}
In \cite{ref8}, the same approach  as in this work has been used, which consists in writing the covariant equations of motion in the momentum reprensatation. A form of the Dirac equation has been put forward that holds for any geometry\footnote{We consider only the geometries which allow a 10-parameter Poincare group, which limits it to the one of constant - positive, negative or vanishing - curvature. See the discussion in \cite{bor2}.} and any choice of the coordinates: 
\begin{equation}\label{flatdirac}
\left(g_{\mu\nu} \tilde \gamma^\mu f^\nu-m\right)\psi=0,
\end{equation}
where 
\begin{equation}
\tilde \gamma^\mu=\gamma^ae_a^\mu=\frac{1}{h}\gamma^\mu-\frac{2h'\alpha^2}{h(h+2h'\alpha^2p^2)}\eta_{ab}\gamma^ap^bp^\mu
\end{equation}
are the projections of ordinary gamma matrices onto the vierbein vector field, and $f^\mu$, called generalized momenta in \cite{ref8}, are the momentum vector fields defined in (\ref{flatgenmom}). The benefit of this form of Dirac's equation is that the objects in it ($\tilde \gamma^\mu$, $f^\mu$) are the true vector fields, meaning they transform accordingly under an isotropic transformation of the coordinates, 
\begin{equation}
p^\mu\to q^\mu= t(\alpha^2p^2)p^\mu \Rightarrow f^\mu(p)\to\tilde f^\mu(q)=\frac{\partial p^\mu}{\partial q^\nu}f^\nu(p),
\end{equation}
with $t$ an arbitrary function, making the covariance of the equation manifest.

Here we write this same equation in a different form: 
\begin{equation}\label{flatdir2}
(\eta_{\mu\nu}\gamma^\mu \tilde p^\nu-m)\psi=0,
\end{equation}
 where

\begin{equation}\label{tildep} 
\tilde p^\mu=d(p,0)\frac{p^\mu}{\sqrt{\eta_{\mu\nu}p^\mu p^\nu}}=h(\alpha^2p^2)p^\mu,
\end{equation}
the $\gamma^\mu$ are the ordinary Dirac matrices, and $\eta_{\mu\nu}$ is the flat metric. Note that all of the objects in the equation are scalars with respect to diffeomorphism transformations of the form (\ref{flatdef}), despite of their appearance (they are still vectors though with respect to Lorentz transformations). This form of the equation is simpler and more practical in calculations, which will become apparent when we calculate Feynman diagrams. Additionaly, this form makes evident that the definition of the Dirac conjugate remains unchanged:

\begin{equation}
\overline \psi = \psi^*\gamma^0,
\end{equation}
as well as

\begin{equation}\label{psib}
\overline\psi(\eta_{\mu\nu}\gamma^\mu \tilde p^\nu+m)=0.
\end{equation}

We point out that the equation (\ref{tildep}) together with (\ref{flatdef}) gives $\tilde p^\mu=\pi^\mu$. 
 This means that the Dirac equation in general coordinates is just a reparametrization of the usual one - this however must be so in our construction, since we demand that the physics depends only on the geometry, and not on the coordinates of the energy-momentum space.

\subsubsection{Free particles and completeness relations}
The solutions of (\ref{flatdirac}) representing (anti)particles with definite energy-momentum $k$ are given as

\begin{equation}\label{psi}
\psi_k(p)=\frac{1}{\sqrt{2d(p',0})}u_p\delta(p\ominus k)
\end{equation}
and 

\begin{equation}
\psi_{-k}(p)=\frac{1}{\sqrt{2d(p',0)}}u_p\delta(p\oplus k),
\end{equation}
where the two solutions satisfy $d^2(0,\pm k)=m^2$, $m$ being the mass of the particle. Here $p'$ is the point obtained from $p$ by setting all but zeroth component to zero, i.e. $p'^\mu=(p^0,0,0,0)$, $u_p$ is the wave amplitude, and the generalised delta-function is defined via

\begin{equation}
\int  f(p) \delta(p\ominus k) d\Omega_p=f(k)
\end{equation} 
for an arbitrary function $f(p)$. 
Inserting (\ref{psi}) into (\ref{flatdir2}) and integrating
, we obtain

\begin{equation}\label{u}
(\eta_{\mu\nu}\gamma^\mu \tilde k^\nu-m)u_k=0
\end{equation}
and 

\begin{equation}
(\eta_{\mu\nu}\gamma^\mu \tilde k^\nu+m)u_{-k}=0.
\end{equation}
From here, one immediately reads out the (anti)fermion propagator, which is the Green function of the Dirac equation,

\begin{equation}
D^f_{\alpha\beta}(k)=\left(\frac{1}{\eta_{\mu\nu}\gamma^\mu \tilde k^\nu\pm m+i\epsilon}\right)_{\alpha\beta}
\end{equation}

Starting from equation (\ref{psib}) and following the same steps as above, we arrive at the corresponding equations for the Dirac adjoints,

\begin{equation}\label{ub}
\overline u_k(\eta_{\mu\nu}\gamma^\mu \tilde k^\nu-m)=0
\end{equation}
and 

\begin{equation}
\overline u_{-k}(\eta_{\mu\nu}\gamma^\mu \tilde k^\nu+m)u_{-k}=0,
\end{equation}
with $\overline u_{\pm k}=u^*_{\pm k}\gamma_0$. We normalise the bispinor amplitudes in the usual way

\begin{align}
\overline u_p u_p&=2m\\
\overline u_{-k} u_{-k}&=-2m,\\
\end{align}
thus reproducing the usual expressions,

\begin{equation}
\overline u_k\gamma^\mu u_k=\overline u_{-k}\gamma^\mu u_{-k}=2\tilde k^\mu,
\end{equation}
up to the replacement $k^\mu\to\tilde k^\mu$. 
%

\subsection{Free gauge fields}
The electromagnetic (EM) tensor in the Cartesian coordinates $\pi$ of the flat energy-momentum space is given as

\begin{equation}
F^{\mu\nu}(\pi)=\pi^\mu A^\nu(\pi)-\pi^\nu A^\mu(\pi).
\end{equation}
The EM tensor in arbitrary coordinates $p$, assuming the diffeomorphism invariance of the theory, is then

\begin{equation}
F^{\mu\nu}(p)=\frac{\partial p^\mu}{\partial \pi^\rho}\frac{\partial p^\nu}{\partial \pi^\sigma}F^{\rho\sigma}(\pi)|_{\pi=hp},
\end{equation}
or
\begin{equation}
F^{\mu\nu}(p)=f^\mu A^\nu(p)-f^\nu A^\mu(p),
\end{equation}
where $f^\mu$ is the generalized momentum vector field, and $A^\mu$ is the vector potential that transforms accordingly under the change of coordinates.
The manifestly invariant form of the free-field action is then

\begin{equation}\label{s}
S=\int g_{\mu\rho}g_{\nu\sigma}F^{\mu\nu}F^{\rho\sigma}d\Omega=\int (d^2(0,p)g_{\mu\nu}(p)-.f_\mu(p) f_\nu(p)) A^\mu(p) A^\nu(p) d\Omega_p.
\end{equation}
In case of the Lorenz gauge, $f_\mu A^\mu=0$, the integrand in the action is a quadratic function of $A^\mu$, so the action is minimal for

\begin{equation}
d^2(0,p) g_{\mu\nu}A^\mu=0,
\end{equation}
which is the Maxwell equation in the absence of sources. From here the photon propagator is readily read out:

\begin{equation}
\tilde D^p_{\mu\nu}(k)=\frac{4\pi}{d^2(k,0)+i\epsilon}g_{\mu\nu}(k).
\end{equation}

\subsection{Interaction and the perturbative expansion}
One usually formulates perturbative QED in the space-time picture: the interaction is defined as a space integral, the covariance being formally recovered via the time-ordered product of an infinite number of interactions. The scattering matrix elements are then represented at each order of the perturbation in terms of the Feynman diagrams, where there is a set of rules, called the Feynman rules, to calculate particular contributions. In practice one always uses the Feynman rules in the momentum representation, because one always scatters particles with definite energy-momenta, not particles with definite start and end points.

We have thus far been very consistent in avoiding any mention of space-time in our construction. We note however that due to the fact that the space-time operators commute for the case of the flat energy-momentum space, the entire analyses described above, where one starts with the space-time representation, and ends up with a set of rules for calculation matrix elements in the momentum reprensetation, is valid also in the case of our generalized energy-momentum coordinates $p$: one can derive everything as in the standard case, and simply "plug-in" the generalized energy-momentum coordinates once the Feynman rules in the momentum representation have been established. As a concrete example, consider the $M$ matrix for the elastic scattering of electrons ($12\to34$); in the standard case, one arrives at

\begin{equation}
M_{fi}=e^2[(\overline u_4\gamma^\mu u_2)D_{\mu\nu}(p_4-p_2)(\overline u_3\gamma^\nu u_1)-(\overline u_4\gamma^\mu u_1)D_{\mu\nu}(p_4-p_1)(\overline u_3\gamma^\nu u_2)],
\end{equation}
where 

\begin{equation}
D^p_{\mu\nu}(k)=\frac{4\pi}{k^2+i\epsilon}\eta_{\mu\nu},
\end{equation}    
is the photon propagator.
The matrix element is manifestly Lorentz invariant. It is than straightforward to write it in the way in which it is also manifestly invariant under the diffeomorphisms:

\begin{equation}\label{M}
\tilde M_{fi}=e^2[(\overline u_4\tilde\gamma^\mu u_2)\tilde D_{\mu\nu}(p_4\ominus p_2)(\overline u_3\tilde\gamma^\nu u_1)-(\overline u_4\tilde\gamma^\mu u_1)\tilde D_{\mu\nu}(p_4\ominus p_1)(\overline u_3\tilde\gamma^\nu u_2)],
\end{equation} 
where $u$ and $\overline u$ are spinor amplitudes satisfying (\ref{u}) and (\ref{ub}), and the photon propagator becomes
\begin{equation}
\tilde D^p_{\mu\nu}(k)=\frac{4\pi}{d^2(k,0)+i\epsilon}g_{\mu\nu}(k).
\end{equation}  
Note that $d^2(p_4\ominus p_2,0)=d^2(p_4,p_2)$, since the addition of momenta is an isometry, as well as

\begin{align}\label{simp}
\tilde \gamma^\mu(k)\tilde\gamma^\nu(k) g_{\mu\nu}(k)&=\gamma^\mu\gamma^\nu \eta_{\mu\nu}\\
\tilde \gamma^\mu(k)\tilde\gamma^\nu(k) f_\mu(k)f_\nu(k)&=\gamma_\mu\gamma_\nu \tilde k^\mu \tilde k^\nu;
\end{align}
the latter fact simplifies expression (\ref{M}), and enables us to use the familiar "traceology" from the standard case, when calculating spin sums and averages, with the understanding that $\cancel p=\gamma_\mu  p^\mu$ from the standard case is to replaced everywhere with  $\tilde{\cancel p}=\gamma_\mu \tilde p^\mu$. This gives finally

\begin{equation}\label{m2}
\begin{split}
\overline {|M|^2}=2e^4& \left[ \frac{1}{t^2}\left(s^2+u^2 -8m^2(s+u) +24m^4       \right)\right.\\
&\left. +\frac{1}{u^2}\left(s^2+t^2 -8m^2(s+t) +24m^4       \right) +\frac{2}{tu}\left(s^2-8m^2s+12m^4     \right)   \right],   
\end{split}
\end{equation}
where the Mandelstam variables are

\begin{align}\label{mand}
s&=d^2(p_1,-p_2)=d^2(p_3,-p_4)\\
t&=d^2(p_1,p_3)=d^2(p_2,p_4)\\
u&=d^2(p_1,p_4)=d^2(p_2,p_3).
\end{align}
This is equal to the standard case, since geodesic distance is invariant on the choice of coordinates.

As convincing as the above argumentation is, for conceptual reasons and in the spirit of our approach, it would be preferable to have a construction that does not rely on space and time, but rather follows enetirely in the energy-momentum space. As it turns out, there exists a formalism of quasi-field operators, or causal fields, developed by several authors \cite{coe,nov,gol}, which allows us presicely that.

\subsection{Quasi-fields}
The motivation for the introduction of quasi-field operators was  that this approach would offer some formal benefits to the theory, in comparison to Feynman's amplitudes approach \cite{fey}. Firstly, one works with the second quantized operators from the onset, not with the wave functions, and secondly, the construction is relativistically invariant from the start. For the details of the theory we refer to the above mentioned papers; here we simply proceed to give its basic elements that are relevant for us. We write everything immediately in the form that is manifestly covariant to the diffeomorphisms.

The fermion quasi-field operators are given by\footnote{We go by the definition in \cite{gol}; in \cite{coe,nov} a slightly different, but equivalent, definition was given.}

\begin{equation}
\psi(p)=\left(g_{\mu\nu}\tilde\gamma^\mu f^\nu-m\right)^{-1/2}\left( \hat a(p)-\hat b^\dagger (-p)  \right)=\left(\tilde{\cancel p}-m\right)^{-1/2}\left( \hat a(p)-\hat b^\dagger (-p)  \right)
\end{equation}
and 

\begin{equation}
\overline\psi(p)=\left( \hat b(p)-\hat a^\dagger (-p)  \right)\left(g_{\mu\nu}\tilde\gamma^\mu f^\nu+m\right)^{-1/2}=\left( \hat b(p)-\hat a^\dagger (-p)  \right)\left(\tilde{\cancel p}+m\right)^{-1/2}
\end{equation}
where the spinor creation and annihilation operators satisfy the following anti-commutation relations

\begin{equation}
\begin{split}
\{\hat a_\alpha(p),\hat a^\dagger_\beta(q)\}&=i\delta_{\alpha\beta}\delta(p\ominus q)\\
\{\hat b_\alpha(p),\hat b^\dagger_\beta(q)\}&=i\delta_{\alpha\beta}\delta(p\ominus q),
\end{split}
\end{equation}
where $\alpha$ and $\beta$ are spinor indices, and all other commutation relations are vanishing. The photon quasi-fields are

\begin{equation}
A^\mu(k)=\left(d^2(0,k)\right)^{-1/2}\left( \hat c^\mu(k)+  \hat c^{\dagger\mu} (-k)   \right),
\end{equation}
where the vector creation and annihilation operators satisfy commutation relations

\begin{equation}
[\hat c^\mu(k),\hat c^{\dagger\nu} (l) ]=ig^{\mu\nu}(k)\delta(k\ominus l).
\end{equation}
From the above, one sees that the operators $\psi(p)$ and $\overline \psi(q)$ anti-commute for any pair of points $p$ and $q$, while the operators $A^\mu(k)$ commute with one another everywhere. The vacuum expectation values of the above defined operators are then 

\begin{equation}\label{vac}
\begin{split}
\langle \psi(p)\overline \psi(q)    \rangle_0&=(\tilde{\cancel p}-m)^{-1}\delta(p\ominus q)\\
\langle A^\mu(k)A^\mu(l)   \rangle_0&=\frac{g^{\mu\nu}(k)}{d^2(k,0)}\delta(k\oplus l),
\end{split}
\end{equation}
with all the remaining vacuum expectation values of products of the pairs of operators vanishing. Finally, the interaction is given by

\begin{equation}\label{int}
\Lambda=e\int d\Omega_p d\Omega_k \overline \psi(p) g_{\mu\nu}(k)\tilde\gamma^\mu(k)A^\nu(k)\psi(p\ominus k).
\end{equation}
The scattering matrix elements are obtained via Green's functions. For our example of electron-electron scattering, these are given by

\begin{equation}\label{K}
K(p_1,p_1; p_3,p_4)=\langle \psi(p_3)\psi(p_4)\hat \sigma \overline\psi(p_1)\overline\psi(p_2)       \rangle_0,
\end{equation}
where the operator 

\begin{equation}\label{sig}
\hat \sigma=e^{i\Lambda}=1+i\Lambda-\frac{1}{2}\Lambda^2+\cdots
\end{equation}
and with obvious generalization for other numbers and types of the incoming and outgoing particles. We note that the interaction operator $\hat \sigma$ commutes with all the quasi-field operators, as well as that Wick's theorem holds.

Using the expansion (\ref{sig}) in (\ref{K}), the first term gives just the sum of products of delta function and inverse propagators, corresponding to no interraction. The second term vanishes, since there is only one photon operator that can not be contracted. The first nontrivial contribution is therefore

\begin{equation}
\begin{split}
K^{(2)}(p_1,p_2; p_3,p_4)=-\frac{e^2}{2}\int d\Omega_p d\Omega_q d\Omega_k d\Omega_l\langle &\psi(p_3)\psi(p_4)\overline\psi(p)\tilde\gamma_\mu(k) A^\mu(k)\psi(p\ominus k)\\
&\overline\psi(q)\tilde\gamma_\mu(l) A^\mu(l)\psi(q\ominus l) \overline\psi(p_1)\overline\psi(p_2)       \rangle_0,
\end{split}
\end{equation}
For instance, in the t-chanel we have, using (\ref{vac}) and the Wick theorem,

\begin{equation}
\begin{split}
K^{(2)}_t(p_1\gamma,p_2\delta; p_3\alpha,p_4\beta)&=-\frac{e^2}{2}(\tilde{\cancel p_3}-m)^{-1}_{\alpha i}(\tilde{\cancel p_4}-m)^{-1}_{\beta m}(\tilde{\cancel p_1}-m)^{-1}_{j\gamma}(\tilde{\cancel p_2}-m)^{-1}_{o\delta}    \\
&\tilde \gamma_{\mu ij}(p_1\ominus p_3)\tilde \gamma_{\nu mo}(p_3\ominus p_1)\frac{g^{\mu\nu}(p_1\ominus p_3)}{d^2(p_1,p_3)}\delta\left( (p_1\ominus p_3)\ominus (p_4\ominus p_2)  \right),       
\end{split}
\end{equation}
where we wrote all the spinor indices explicitly, and the summation over the repeated ones is implicit. Using (\ref{simp}) together with the fact that $\tilde \gamma^\mu(k)=\tilde \gamma^\mu(-k)$, as well as

\begin{equation}
(\tilde{\cancel p}-m)^{-1}=\frac{\tilde{\cancel p}+m}{d^2(p,0)-m^2},
\end{equation}
we finally get
\begin{equation}
\begin{split}
K^{(2)}_t(p_1\gamma,p_2\delta; p_3\alpha,p_4\beta)&=-\frac{e^2}{2}\frac{[(\tilde{\cancel p_3}+m)\gamma^\mu(\tilde{\cancel p_1}+m)]_{\alpha\gamma}[(\tilde{\cancel p_4}+m)\gamma^\nu(\tilde{\cancel p_2}+m)]_{\beta\delta}}{(p_1^2-m^2)(p_2^2-m^2)(p_3^2-m^2)(p_4^2-m^2)}   \\
&\frac{\eta_{\mu\nu}}{d^2(p_1,p_3)}\delta\left( (p_1\ominus p_3)\ominus (p_4\ominus p_2)  \right),       
\end{split}
\end{equation}
with a similar expresion for the u-chanel. Identifying, as in the standard case, the scaterring matrix elements with the poles of the Green function once the external particles are put on shell \cite{lsz}, we recognise the familiar traces appearing in the spin averaged square of the amplitude and recover the result (\ref{m2}).

\section{The adS case}
The adS geometry of the energy-momentum space corresponds to the Snyder model, which was the first proposed model of noncommutativity. On the level of Heisenberg algebra, it is characterised by the commutation relations

\begin{equation}
[\hat x_\mu, \hat x_\nu]=\beta^2J_{\mu\nu}, \ \ \ \ \ \ [\hat p^\mu,\hat p^\nu]=0, \ \ \ \ \ \ [\hat x_\mu, \hat p^\nu]=f_1(\beta^2p^2)\delta_\mu^\nu+f_2(\beta^2p^2)\eta_{\mu\rho}p^\rho p^\nu,
\end{equation}
where again the closure af algebra constrains the functions $f_1$ and $f_2$ \cite{bor}. We shall not consider the algebra, but continue with our approach through the energy-momentum space. Having spent so much time in the last section to establish what amounts to just a reparametrization of the ordinary theory, we are now in a position to simply apply our findings to the case that is really of our interest, that of the AdS geometry. Our task reduces to just finding the geometry elements that are relevant to us.

\subsection{Elements of the geometry}

An adS space is realised as a surface embedded in 4+1 dimensional background,

\begin{equation}\label{emb}
\pi^2-\pi_4^2=-1/\beta^2
\end{equation}
with physical momenta defined through an arbitrary isotropic projection from the "northern hemisphere"

\begin{equation}
p^\mu=g(\beta^2p^2)\pi^\mu, \ \ \ \ \ \ \pi^\mu=h(\beta^2p^2)p^\mu, \ \ \ \ \pi_4=\frac{1}{\beta}\sqrt{1+h^2\beta^2p^2}.
\end{equation}
The metric and its inverse are given by 
\begin{align}\label{met}
g_{\mu\nu}= &h^2\eta_{\mu\nu}+\beta^2\frac{4h'(h+h'\beta^2p^2)-h^4}{1+h^2\beta^2p^2}\eta_{\mu\rho}p^\rho\eta_{\nu\sigma}p^\sigma \\
g^{\mu\nu}=&\frac{1}{h^2}\eta^{\mu\nu}-\beta^2\frac{4h'(h+h'\beta^2p^2)-h^4}{h^2(h+2h'\beta^2p^2)}p^\mu p^\nu,
\end{align}
where
\begin{equation}
h'\equiv \frac{\partial h}{\partial(\beta^2p^2)}, \ \ \ \ g'\equiv\frac{\partial g}{\partial(\beta^2\pi^2)}. 
\end{equation}
The determinant of the metric and the volume element are 
\begin{equation}\label{vol}
\text{det}g=-\frac{h^6(h+2h'\beta^2p^2)^2}{1+h^2\beta^2p^2}, \ \ \ \ \ \ \ d\Omega_p=\frac{h^3(h+2h'\beta^2p^2)}{\sqrt{1+h^2\beta^2p^2}}d^4p
\end{equation}
The rule of the momenta addition is 
\begin{equation}
(p\oplus q)^\mu=g(\alpha^2\kappa^2)\kappa^\mu, \ \ \ \ \ \kappa^\mu=h_qq^\mu+ \left(\sqrt{1+h_q^2\beta^2q^2}+\frac{\beta^2h_ph_q(pq)}{1+\sqrt{1+h_p^2\beta^2p^2}}       \right) h_pp^\mu,
\end{equation}
where $h_p=h(\beta^2p^2)$, $h_q=h(\beta^2q^2)$ and $(pq)=\eta_{\mu\nu}p^\mu q^\nu$. \\
The distance function is 
\begin{equation}\label{dist}
d(p,q)=\frac{1}{\beta}\text{Arcosh}\left( h_ph_q\beta^2(pq)-\sqrt{1+h_p^2\beta^2p^2}\sqrt{1+h_q^2\beta^2q^2}    \right).
\end{equation}
The generalized momenta are 

\begin{equation}\label{genf}
f_\mu\equiv\frac{1}{2}\frac{\partial d^2(p,0)}{\partial p^\mu}=\frac{h+2h'\beta^2p^2}{\sqrt{1+h^2\beta^2p^2}}\frac{d(p,0)}{\sqrt{p^2}}\eta_{\mu\rho}p^\rho, \ \ \ \ f^\mu=g^{\mu\nu}f_\nu=\frac{1+h^2\beta^2p^2}{\sqrt{1+h^2\beta^2p^2}(h+2h'\beta^2p^2)}\frac{d(p,0)}{\sqrt{p^2}}p^\mu
\end{equation}
The vierbein is 
\begin{equation}\label{vie}
e^\mu_a\equiv \lim_{q\to 0} \delta_a^\nu \frac{\partial (p\oplus q)^\mu}{\partial q^\nu}=\frac{1}{h}\delta_a^\mu + \left( \frac{\beta^2h}{1+\sqrt{1+h^2\beta^2p^2}}- \frac{2h'\beta^2\sqrt{1+h^2\beta^2p^2}}{h(h+2h'\beta^2p^2)}   \right)p_ap^\mu
\end{equation}

\subsection{The dynamics}
The free Dirac equation is again given by\footnote{In order not to clog our expressions, we use the same notation as in the flat case - no confusion should arise due to this.}
\begin{equation}\label{adsdirac}
\left(g_{\mu\nu} \tilde \gamma^\mu f^\nu-m\right)\psi=0,
\end{equation}
where now
\begin{equation}
\tilde \gamma^\mu=\gamma^ae_a^\mu=\frac{1}{h}\gamma^\mu+ \left( \frac{\beta^2h}{1+\sqrt{1+h^2\beta^2p^2}}- \frac{2h'\beta^2\sqrt{1+h^2\beta^2p^2}}{h(h+2h'\beta^2p^2)}   \right)\eta_{ab}\gamma^ap^bp^\mu
\end{equation}
are the projections of ordinary gamma matrices onto the vierbein vector field (\ref{vie}), and $f^\mu$, are the momentum vector fields defined in (\ref{genf}). As in the flat case, the Dirac equation can be put in a simpler form
\begin{equation}
(\eta_{\mu\nu}\gamma^\mu \tilde p^\nu-m)\psi=0,
\end{equation}
 where again

\begin{equation}\label{tildepads} 
\tilde p^\mu=d(p,0)\frac{p^\mu}{\sqrt{\eta_{\mu\nu}p^\mu p^\nu}},
\end{equation}
with the understanding that the distance function in (\ref{dist}) is now to be used.
What concerns the gauge boson part, we again have
\begin{equation}
F^{\mu\nu}(p)=f^\mu A^\nu(p)-f^\nu A^\mu(p),
\end{equation}
where $f^\mu$ is the momentum vector field (\ref{genf}), and $A^\mu$ is the vector potential that transforms accordingly under the change of coordinates.
The manifestly invariant form of the free-field action is then

\begin{equation}
S=\int g_{\mu\rho}g_{\nu\sigma}F^{\mu\nu}F^{\rho\sigma}d\Omega,
\end{equation}
with the metric and volume element (\ref{met}) and (\ref{vol}). 

Every single step that led to the amplitude (\ref{m2}) for our paradigmatic process of electron-electron scattering, both through the wave amplitudes and via the quasi-field operators, holds in the case of adS geometry. Thus the result (\ref{m2}) holds here as well, with the understanding that the distance function (\ref{dist}) is to be used for the Mandelstam variables in (\ref{mand}). For other processes, one can simply use the modified Feynman rules, with the obvious generalizations for the vertex factors, propagators, delta functions and integrations.

\subsection{Conclusion and outlook}
In the algebraic approach to the noncummutative models, one works in the position reprensetation and uses the deformed commutation algebra (3), (66) as the starting point. These are then used in the construction of star products between functions, which establishes the duality between the algebra of functions of commutative and noncommutative coordinates. This approach however leads to considerable difficulties both on the technical as well as the  conceptual level. On the technical level, there is the operator ordering problem, as well as the problem of momenta addition for a general realization (the co-product rule). On the conceptual level, there is the problem of defining fundamental dynamical quantities, such as Hamiltonian or Lagrangian. 

 In the approach presented in this paper, these issues are either simply resolved or do not appear at all. The key novelty - the demand of the diffeomorphism symmetry on the momentum space - resolves the conceptual issues, since it provides a principle from which the dynamic quantities follow. On the technical level, one avoids the ordering problem by working in the momentum representation, while the momenta addition rule follows simply from the underlying geometry.

Even though the main motivation for the study of noncommutative models nowdays stems from the quantum gravity research, originally Snyder proposed his model to cure the divergencies appearing in the field theory. One immediately notices that due to the finitness of the volume of the adS space,

\begin{equation}
\int d\Omega_p=\frac{4}{3}\frac{\pi^2}{\beta^4}
\end{equation}
all the integrals that appear in calculations of the amplitudes of various processes beyond the tree-level will be finite. This is manifest from the fact that the Green functions containing the loop integrals will always consist of rational functions integrated over a finite volume of the momentum space. Even though the renormalization of the standard QED with the help of dimensional regularisation works well in reproducing, to a great accuracy, the results of our scattering experiments, from mathematical point of view the situation is far from satisfying - the bare masses and charges are still infinite. The proposal to depart from the flat geometry of the energy-momentum space seems much more elegant in this respect. The only price one pays is the appearance of an additional dimensional quantity in the theory. This additional constant, however, need not be a new independent constant of nature. As proposed in \cite{mark2}, it can be identified with the Planck's length, i.e. be built from the know physical constants $\hbar$, $c$ and $G$. In this case, the minimal length is identified with the phenomenological minimal length, that is the Schwarzschild radius of a black hole that appears when a large amount of energy is squeezed into a small space \cite{me,mark1}.

An additional feature of the adS geometry of the energy-momentum space, not discussed in the literature, is that it allows for an alternative description of antiparticles. Instead of their usual interpretation as the states with negative energy, they could retain a strictly positive energy, but can have the value of the parameter $\pi_4$ from (\ref{emb}) negative instead. In this case the vacuum would not be unique, but degenerate, with particles belonging to northern half with the vacuum at $\pi_4=1/\beta$, and the antiparticles to the southern half with $\pi_4=-1/\beta$. In this cases, only strictly massless particles could be their own antiparticles. This possibility deserves a further study.

Finally, an importaint question of the gauge symmetry has not been touched upon. The action (\ref{s}) is obviously invariant under the transformation 

\begin{equation}
A^\mu(p)\to A^\mu(p)+\Lambda(\beta^2p^2)p^\mu
\end{equation}
for any function $\Lambda$. It remains to determine the accompanying transformation on $\psi(p)$ that leaves (\ref{int}) invariant. In particular, the relation between gauge invariance and diffeomorphism invariance should be investigated. This is the topic of the ongoing study.

\section*{Acknowledgment}
I am grateful to S.A. Franchino-Vi\~nas for the stimulating correspondence in the early stages of the work.

\end{document}